\documentclass[twocolumn]{aastex631}
\usepackage{graphicx}

\begin{document}

\title{Inspecting Cloudy Substellar Atmospheres with JWST MIRI Synthetic Magnitudes from Spitzer Mid-infrared Spectra}

\author[0009-0009-3024-5846]{Jolie L'Heureux}
\affiliation{Department of Physics, Graduate Center, City University of New York, 365 5th Ave., New York, NY 10016, USA}
\affiliation{Department of Astronomy, Columbia University, New York, NY 10027, USA}

\author[0000-0002-2011-4924]{Genaro Su\'arez}
\affiliation{Department of Astrophysics, American Museum of Natural History, Central Park West at 79th Street, NY 10024, USA}

\author[0000-0003-0489-1528]{Johanna M. Vos}
\affiliation{School of Physics, Trinity College Dublin, The University of Dublin, Dublin 2, Ireland}

\author[0000-0003-3050-8203]{Stanimir Metchev}
\affiliation{Department of Physics and Astronomy, Western University, 1151 Richmond St, London, Ontario N6A 3K7, Canada}
\affiliation{Institute for Earth and Space Exploration, Western University, 1151 Richmond St, London, Ontario N6A 3K7, Canada}

\author[0000-0001-6251-0573]{Jacqueline K. Faherty}
\affiliation{Department of Astrophysics, American Museum of Natural History, Central Park West at 79th Street, NY 10024, USA}

\author[0000-0003-0548-0093]{Sherelyn Alejandro Merchan}
\affiliation{Department of Astrophysics, American Museum of Natural History, Central Park West at 79th Street, NY 10024, USA}
\affiliation{Department of Physics, Graduate Center, City University of New York, 365 5th Ave., New York, NY 10016, USA}

\author[0000-0002-1821-0650]{Kelle L. Cruz}
\affiliation{Department of Astrophysics, American Museum of Natural History, Central Park West at 79th Street, NY 10024, USA}
\affiliation{Department of Physics and Astronomy, Hunter College, City University of New York, 695 Park Avenue, New York, NY 10065, USA}
\affiliation{Department of Physics, Graduate Center, City University of New York, 365 5th Ave., New York, NY 10016, USA}

\begin{abstract}

We examine the positions of substellar objects in mid-infrared color-magnitude and color-color diagrams to distinguish between cloudy and cloud-free atmospheres. 
Using Spitzer mid-infrared spectra of 113 M5–T9 ultracool dwarfs, we derive synthetic photometry for the JWST MIRI F560W, F770W, F1000W, and F1280W filters, which cover key absorption features including the $\sim$9 $\mu$m silicate signature. 
We find that diagrams involving F770W and F1000W best separate L-type objects with silicate clouds in their photospheres. 
L dwarfs with $m_{\rm{F770W}}-m_{\rm{F1000W}}<0.03$~mag are seven times more likely to host cloudy atmospheres. 
Diagrams using F1000W and F1280W are less informative due to the lower signal of the spectra at long wavelengths. 

Current model predictions struggle to reproduce the positions of cloudy, warm brown dwarfs, likely because atmospheric models underestimate the $\sim$9~$\mu$m silicate feature. Cloudy Sonora Diamondback models better match the observed trends, 
although this may reflect improvements capturing indirect effects of clouds on the 6.25~$\mu$m water absorption feature rather than accurately modeling the silicate feature itself.
Our analysis indicates that JWST MIRI photometry can efficiently identify new cloudy extrasolar atmospheres for targeted spectroscopic follow-up, optimizing the use of telescope time.

\end{abstract}



\section{Introduction}
\label{sec:intro}

Clouds are an important component of planetary atmospheres and are common in extrasolar atmospheres from exoplanets to brown dwarfs
\citep{Gao_2021}. They influence most aspects of planetary atmospheres such as radiation transport and atmospheric chemistry \citep{Barstow_2016}. Therefore, a better understanding of these clouds such as their formation, evolution, chemical composition, and physical properties is crucial for understanding planetary atmospheres as a whole including their thermal structure, composition, and habitability \citep{Marley_2013}.

While \citet{Cushing_2006} reported the first unequivocal direct detections of cloudy atmospheres in two brown dwarfs, 
\citet{Suarez_2022} found that silicate clouds are common in brown dwarf atmospheres with effective temperatures of roughly 2000–1300 K (corresponding to $\approx$ L2--L8). 
This conclusion was based on the detection of a $\sim$ 9 $\mu$m silicate absorption feature in L dwarfs---the most direct observational evidence of silicate clouds in atmospheres---using data from the now retired NASA's Spitzer Space Telescope \citep{Werner_2004} obtained with the Infrared Spectrograph \citep[IRS,][]{Houck_2004}. 
The same dataset also showed that silicate clouds are more abundant in the atmospheres of young objects and at equatorial latitudes 
\citep{Suarez_2023a,Suarez_2023b}. 

Mid-Infrared wavelengths, which span $\sim$ 5--30 $\mu$m, sample prominent absorption features of water, methane, ammonia, and silicate in ultracool dwarfs \citep{Roellig_2004, Cushing_2006, Looper_2008}, making mid-infrared spectra essential for studying silicate clouds. 
To date, mid-infrared spectra have been obtained mainly with Spitzer IRS, with just over $\sim$100 substellar objects observed \citep[][and references therein]{Suarez_2022}. A growing sample of mid-infrared spectra is becoming available with the JWST Mid-Infrared Instrument \citep[MIRI;][]{Wells_2015}, with $\sim$50 spectra of L-type dwarfs obtained so far, including 16 from Survey 3930 (PI: Metchev; Barber et al. 2026, in prep.), 5 from GO 3486 (PI: Vos; Lam et al. 2026, submitted), 9 from GO 4668 (PI: Burgasser) using the MIRI low-resolution spectrometer (LRS) and 1 from \citet{Miles_etal2023}, 2 from GO 2288 (PI: Lothringer), 9 from GO 5474 (PI: Su\'arez), and 25 to be acquired in GO 11448 (PI: Faherty) using the MIRI medium resolution spectrometer (MRS).

In this paper, we propose an alternative method to detect and study general trends of cloudy substellar atmospheres using mid-infrared photometry from JWST MIRI. The MIRI filters have full wavelength coverage of the silicate absorption and the continuum regions around the feature, making them well-suited for identifying silicate clouds. 
Pre-electing cloudy targets via photometry provides an efficient way to optimize follow up spectroscopy of new cloudy atmospheres for further characterization.
Because acquiring photometry is much less time consuming than spectra, JWST MIRI photometry presents an avenue to directly detect silicate clouds in extrasolar atmospheres. 

This paper presents mid-infrared color-magnitude diagrams (CMDs) and color-color diagrams (CCDs) to explore the detectability of substellar objects with silicate clouds in their photospheres by producing synthetic JWST MIRI photometry from Spitzer/IRS spectra. The paper is organized as follows. An overview of the Spitzer IRS mid-infrared spectra and JWST MIRI mid-infrared photometry used in this study is given in Section \ref{sec:data}. Section~\ref{sec:methods} explains how we derive synthetic photometry. In Section \ref{sec:results} we construct CMDs and CCDs using synthetic and observed MIRI magnitudes and place them in context with model  predictions. 
We conclude and summarize this study in Section~\ref{sec:summary}.

\section{Data}
\label{sec:data}

\subsection{Mid-Infrared Spitzer IRS Spectra}
\label{sec:spitzer_spectra}
\citet{Suarez_2022} reprocessed the spectra of all field  ultracool objects observed with the Spitzer IRS. The sample contains 113 M5–T9 dwarfs, including 12 late-M dwarfs, 69 L dwarfs, and 32 T dwarfs. All spectra have a low-resolution of $R\approx$100 and 
the majority have a wavelength coverage of 5.2--14.2 $\mu$m. Only the five latest T dwarfs have a shorter wavelength coverage of  7.5--14.2 $\mu$m and 18 of the spectra
have a broader coverage of 5.2--38 $\mu$m. The median signal-to-noise (S/N) ratio of the spectra is 
$\approx$20 at 6 $\mu$m and $\approx$ 10 at 12 $\mu$m. 

Although there are a few moderately young (a few tens or hundreds of Myr) objects in the Spitzer IRS sample, most of them are old field ($\gtrsim$ 1 Gyr) brown dwarfs. We use the same spectral type and denotation considerations as in \citet{Suarez_2022}: 
optical spectral type for M and L dwarfs and infrared spectral type for T dwarfs. 
The target names are abbreviated as hhmm$\pm$ddmm. This format specifies the right ascension (hours and minutes) and declination (degrees and minutes) at the J2000.0 equinox. Table~1 in \citet{Suarez_2022} lists the 113 mid-M to late-T dwarfs with Spitzer IRS spectra and their spectral types.

\subsection{Mid-Infrared JWST MIRI Photometry}
\label{sec:jwst_spectra}

The mid infrared photometry that we used was acquired as part of a JWST Cycle 1 GO 2124\footnote{\url{https://www.stsci.edu/jwst/science-execution/program-information?id=2124}} program (PI: Faherty). The program obtained NIRSpec near-infrared spectra and MIRI mid-infrared photometry for 12 cold brown dwarfs including late-T and early-Y dwarfs. The photometry was obtained using the MIRI filters F1000W, F1280W, and F1800W, which are centered at 10 $\mu$m, 12.8 $\mu$m, and 18 $\mu$m, respectively. We used the photometry automatically reduced by the JWST pipeline available via MAST (Mikulski Archive for Space Telescopes) and considered the aperture-corrected total magnitudes measured in the Vega magnitude system.

\section{Methods: Compute Synthetic Photometry}
\label{sec:methods}

\subsection{Selection of JWST MIRI Filters}
\begin{figure}[ht]
    \includegraphics[width=1\linewidth]{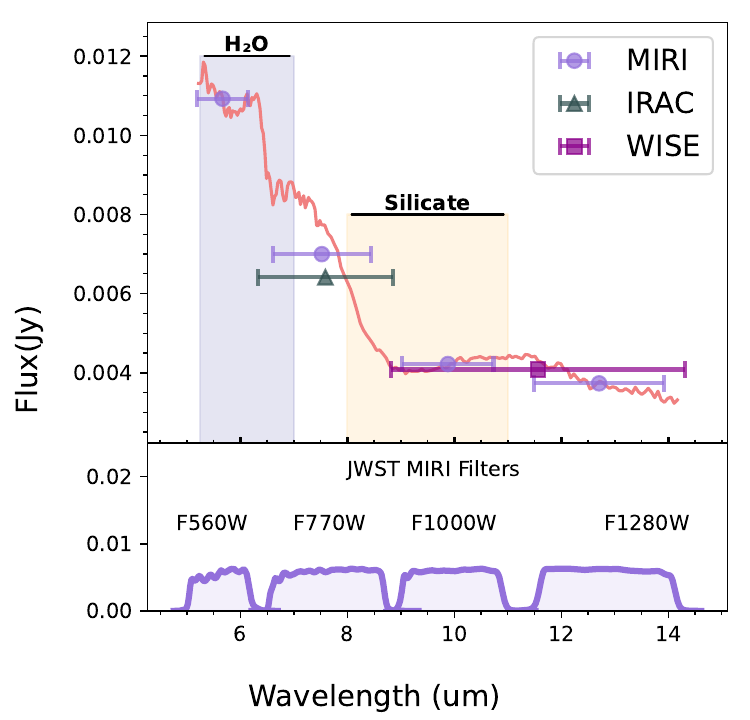}
    \caption{Spitzer IRS spectrum of 2148+4003 (red curve) and photometry from IRAC Ch4 (green), WISE W3 (dark purple), and synthetic fluxes using the diverse MIRI (light purple) filters in the bottom panel. 
    The main spectral features in the spectrum are indicated. 
    The F560W filter covers the water absorption and the F770W, F1000W, and F1280W filters are centered, respectively, before, within, and after the silicate absorption signature in the spectrum. 
    The horizontal bars for photometric measurements correspond to the effective wavelength of each filter.} 
    \label{fig:compare_filters}
\end{figure}

To derive synthetic photometry from all the 113 Spitzer IRS spectra of ultracool objects, we considered the F560W (4.9--6.4 $\mu$m), F770W (6.5--8.8 $\mu$m), F1000W (8.8--11.1 $\mu$m), and F1280W (11.3--14.3 $\mu$m) filters from JWST MIRI. 
As shown in Figure~\ref{fig:compare_filters}, these filters cover the 5.2--14.2~$\mu$m wavelength range of most Spitzer IRS spectra. 
The F560W filter probes the 6.25~$\mu$m water absorption feature present in all the IRS spectra. For L-type objects, the F1000W filter probes the 8--11~$\mu$m silicate absorptions, while F770W samples the pseudo-continuum at shorter wavelengths and F1280W sample the pseudo-continuum at longer wavelengths on either side of the feature. For cooler T-type dwarfs, F770W probes the 7--9~$\mu$m methane absorption feature, and F1000W covers the 10--11.2~$\mu$m ammonia absorption.

Figure~\ref{fig:compare_filters} also shows IRAC Ch4 and WISE W3 photometry. However, these filters are significantly broader than the MIRI bands and therefore are less effective for isolating the effect of silicate clouds. In particular, the WISE W3 filter spans both the silicate absorption and the adjacent pseudo-continuum, making it difficult to separate the contribution of this absorption feature. 
This contrast highlights the advantage of MIRI photometry to probe silicate absorption and efficiently identify cloudy atmospheres. 

\subsection{Extending Spectra to Cover All Filters}
\label{sec:spectral_extension}
We note that the Spitzer spectra do not fully cover the F560W filter, as the shortest wavelength in the data is 5.2~$\mu$m while the filter transmission extends down to 4.9~$\mu$m.
We estimated that $\approx$12\% of the F560W filter transmission is out of the spectra wavelength coverage. To correct for this, we appended 4.9–5.2 $\mu$m model spectra to the Spitzer spectra using ATMO 2020 atmospheric models \citep{Phillips_2020}.

For this purpose, we considered model spectra with parameters appropriate for our sample and used models spanning a range of parameters for each target to estimate the flux uncertainties associated with the appended model spectra. Specifically, we adopted: $i)$ the effective temperatures corresponding to the spectral types of the objects according to the relation in \citet{Filippazzo_2015}, $ii)$ surface gravity of $\log g = 5.0$, which is suitable for our sample consisting mainly of old field objects \citep{Suarez_2022}, and $iii)$ the model grid considering equilibrium chemistry.
The ATMO 2020 models assume solar metallicity. 
For each target, we normalized the selected model spectra to the median flux of the Spitzer spectrum between 5.2 and 5.5 $\mu$m and then attached the models to the Spitzer spectra to have full spectral coverage of the F560W filter.

To quantify the uncertainties introduced by the choice of atmospheric parameters, we explored a subset of model spectra spanning $\Delta T_{\mathrm{eff}} = \pm 100$ K around the temperature inferred from the spectral type, surface gravities of $\log g = 4.0-5.5$, and models including disequilibrium chemistry due to vertical mixing with eddy diffusion coefficient of $\log K_{zz}=4$ (weak mixing) and $\log K_{zz}=6$ (strong mixing), in addition to the grid assuming equilibrium chemistry ($\log K_{zz}=0$).
For each target, we constructed appended spectra using all combinations of these parameters and evaluated the resulting spread in flux across the appended spectral region (4.9–5.2 $\mu$m). The flux uncertainties assigned to the appended portion of the spectrum were taken as the maximum deviation among all model combinations at each wavelength. This approach ensures that the error bars account simultaneously for uncertainties in effective temperature, surface gravity, and vertical mixing efficiency.

\subsection{Synthetic Photometry}
The synthetic photometry was calculated over a bandpass ($\lambda_{A}, \lambda_{B}$) by convolving the Spitzer fluxes ($F_{\lambda}$) with the MIRI filters ($S_{\lambda}$):

\begin{equation}
F_{syn} = \frac{\int_{\lambda_{A}}^{\lambda_{B}}F_{\lambda}S_{\lambda}d\lambda}{\int_{\lambda_{A}}^{\lambda_{B}}
S_{\lambda}d\lambda}
\label{eq:syn_flux}
\end{equation}

We associate synthetic flux errors ($eF_{syn}$) by keeping the fractional errors of the spectra fluxes, including the flux uncertainties from model extensions for F560W.

We converted the synthetic fluxes into magnitudes using the zero point flux for each filter ($F_0$) from the Spanish Virtual Observatory (SVO)\footnote{\url{http://svo2.cab.inta-csic.es/theory/fps/}}:

\begin{equation}
m_{syn} = -2.5 * log_{10}(F_{syn}/F_{0})
\end{equation}

The magnitude errors were obtained by propagating the synthetic flux uncertainties.

We transformed the apparent magnitudes $m_{syn}$ into absolute magnitudes $M_{syn}$ considering the distance modulus $m_{syn}-M_{syn}$ from parallaxes in the UltracoolSheet catalog
\citep{Best_etal2025} and using $d$=1/parallax:

\begin{equation}
m_{syn}-M_{syn} = 5\log_{10} d - 5
\end{equation}


\section{Results}
\label{sec:results}
\begin{figure*}[ht]
    \centering
    \includegraphics[width=1\linewidth]{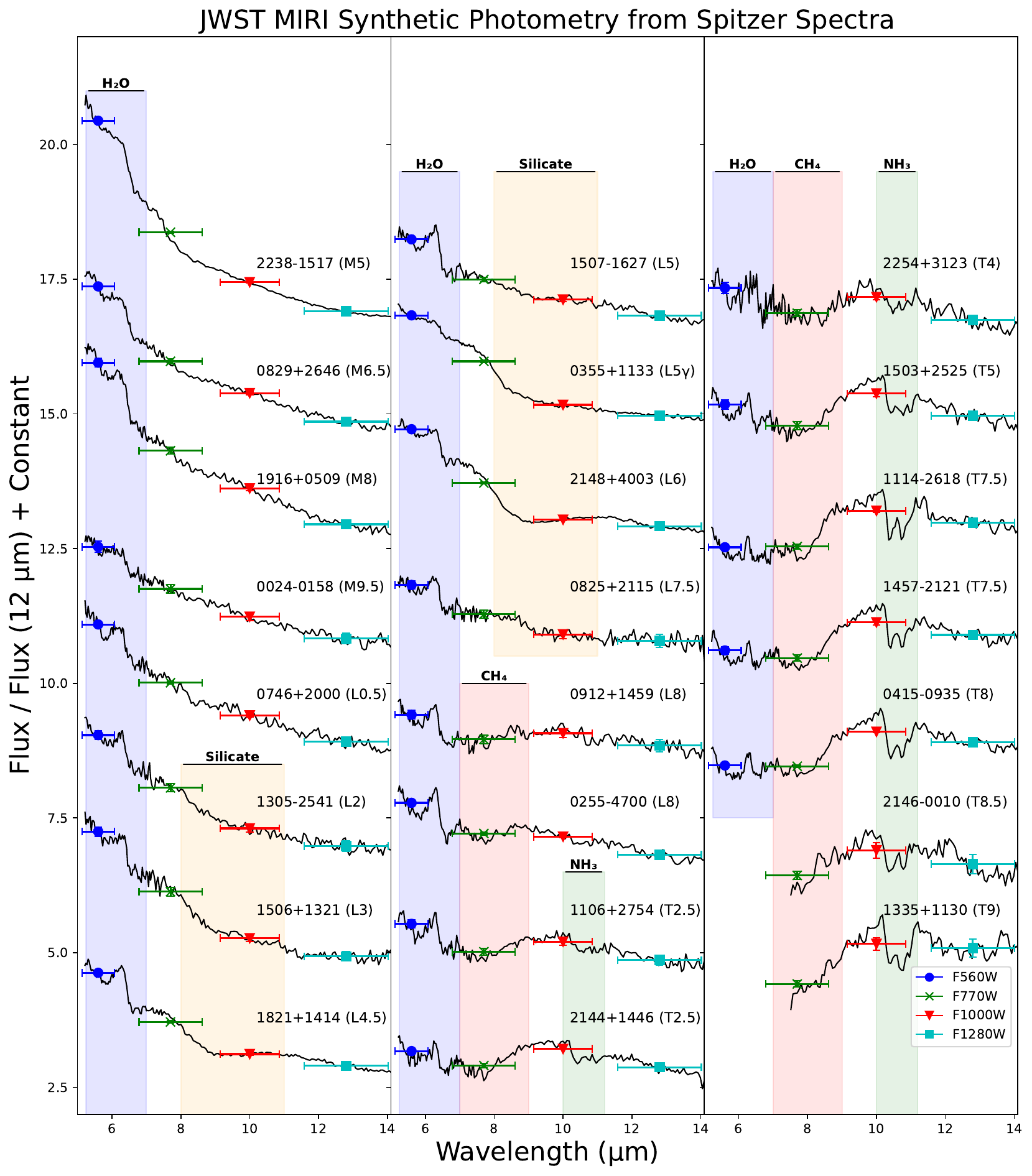}
	\caption{Subset of 23 of the 113 Spitzer IRS spectra analyzed in this study for late-M, L, and T dwarfs. Our derived MIRI synthetic fluxes using the F560W, F770W, F1000W, and F1280W filters are overplotted, as indicated in the label. The subset contains four M, 10 L, and 9 T-type dwarfs.  
    Key molecular signatures are indicated, namely water (H$_2$O), silicates, methane (CH$_4$), and ammonia (NH$_3$), which are present, respectively, in M5--T9, L1--L8, $\gtrsim$L8, and $\gtrsim$T2.5 spectra \citep{Suarez_2022}. }
    \label{fig:plot_subset}
\end{figure*}

We obtained MIRI F560W, F770W, F100W, and F1280W synthetic photometry for the 113 M5--T9 Spitzer IRS spectra in \citet{Suarez_2022}. 
Figure~\ref{fig:plot_subset} shows the synthetic fluxes for a representative subset of these spectra that spans the full range of spectral types. 
We used the synthetic magnitudes to create CMDs and CCDs to investigate the position of cloudy and cloud-free atmospheres and the evolutionary sequences from the warmest to the coldest brown dwarfs. 

We note that the detection or lack of the silicate absorption feature in the data provides direct evidence for the presence or absence of silicate clouds in the mid-infrared photosphere \citep[e.g.,][]{Cushing_2006,Looper_2008}. Accordingly, unless otherwise specified, references to cloudy or cloud-free atmospheres refer specifically to whether silicate clouds are present or absent in the atmospheric layers probed at mid-infrared wavelengths i.e., the mid-infrared photosphere.

\subsection{Synthetic Color-Magnitude Diagrams}
\label{sec:syn_color_mag}
We created five CMDs: 
$M_{\rm{F770W}}$ vs. $m_{\rm{F770W}}-m_{\rm{F1000W}}$, $M_{\rm{F1000W}}$ vs. $m_{\rm{F770W}}-m_{\rm{F1000W}}$, $M_{\rm{F1000W}}$ vs. $m_{\rm{F1000W}}-m_{\rm{F1280W}}$, $M_{\rm{F1280W}}$ vs. $m_{\rm{F1000W}}-m_{\rm{F1280W}}$,
and $M_{\rm{F770W}}$ vs. $m_{\rm{F560W}}-m_{\rm{F770W}}$.
Figure~\ref{fig:CMDs} shows the former four diagrams, and Figure~\ref{fig:CMD2} displays the latter. 
We use these diagrams to identify general trends in the data and to compare them with theoretical predictions.

\subsubsection{Empirical Trends}
\label{sec:empirical_trends}

To examine the position of objects with cloudy and cloud-free atmospheres in the CMDs in Figure~\ref{fig:CMDs} and 
Figure~\ref{fig:CMD2}, we identified a subset of objects with silicate clouds in their visible atmospheres based on strong silicate absorption in the Spitzer IRS spectra, as follows.

\citet{Suarez_2022,Suarez_2023a} introduced a silicate index definition as a measurement of the silicate absorption depth. Briefly, the silicate index compares the flux of an interpolated continuum at the absorption location to the median flux within the feature, where higher index values indicate stronger silicate absorptions. 
We consider as cloudy atmospheres L-type dwarfs with a silicate index 0.1 higher than the corresponding median index for objects with the same spectral type, as in \citet{Suarez_2023a}.
This subset of cloudy atmospheres is highlighted in Figures~\ref{fig:CMDs} and \ref{fig:CMD2}.

In all the CMDs in Figures~\ref{fig:CMDs} we observe that the subset of objects with cloudy atmospheres roughly separate from other objects with similar spectral types. 
Cloudy atmospheres preferentially have bluer $m_{\rm{F770W}}-m_{\rm{F1000W}}$ colors and redder $m_{\rm{F1000W}}-m_{\rm{F1280W}}$ colors. This occurs because the F1000W filter is very sensitive to silicate absorption, so it reflects the silicate absorption diversity reported in \citet{Suarez_2022}. 
Objects with strong silicate absorption due to cloudy atmospheres have fainter $m_{\rm{F1000W}}$ magnitudes and, therefore, bluer $m_{\rm{F770W}}-m_{\rm{F1000W}}$ colors and redder $m_{\rm{F1000W}}-m_{\rm{F1280W}}$ colors. 
Contrary, the $m_{\rm{F560W}}-m_{\rm{F770W}}$ color in Figure~\ref{fig:CMD2} is less affected by silicate clouds as the F560W and F770W filter transmissions center before the silicate absorption. Therefore, the subset of cloudy atmospheres does not clearly separate from the rest of the sample in the CMD involving the F560W and F770W filters.

Although most cloudy atmospheres occupy a different region in the CMDs in Figure~\ref{fig:CMDs}, the uncertainties in the diagrams involving the F1280W filter (bottom panels) are larger due to the noisier Spitzer spectra at long wavelengths (Section~\ref{sec:spitzer_spectra}).
Therefore, the CMDs considering the F770W and F1000W filters (top panels) provide a more robust separation between cloudy and cloudless atmospheres. 
In the $M_{\rm{F770W}}$ vs. $m_{\rm{F770W}}-m_{\rm{F1000W}}$ CMD (top left panel in Fig. \ref{fig:CMDs}), 
we drew three boxes to show where the M, L, and T dwarfs mainly fall on the diagram. The highest box in purple includes 92\% (11 of 12) of the late-M dwarfs in the sample.
The middle box in green encompasses 91\% (63 of 69) of the L dwarfs. 
However, the region with late-M and L dwarfs slightly overlaps.
There are 9 M dwarfs in the L dwarf box, while 5 L dwarfs lie within the M dwarf region.
The bottom box in red contains all the 32 T dwarfs in our sample.
Nevertheless, two L dwarfs fall in the T dwarf box. 
These boxes provide a quick visualization of where M, L, and T dwarfs predominantly fall within the diagram, complementing the auxiliary color scale. 

Focusing on the L dwarfs in the top left panel in Figure~\ref{fig:CMDs} and considering only objects with $m_{\rm{F770W}}$ and $m_{\rm{F1000W}}$ more precise than 1\%, 
we found that 70\% (7 of 10) of the cloudy atmospheres have $m_{\rm{F770W}}-m_{\rm{F1000W}}$ colors bluer than 0.03~mag, while only 3\% (1 of 36) of the cloudless atmospheres have similarly blue colors. 
Thus, 7 of the 8 (88\%) L dwarfs with $m_{\rm{F770W}}-m_{\rm{F1000W}}<0.03$~mag are cloudy objects. 
This indicates that L dwarfs (or dwarfs with $M_{\rm{F770W}}\approx$9.2--10.5~mag) with $m_{\rm{F770W}}-m_{\rm{F1000W}}<0.03$~mag are seven times more likely to have cloudy than cloudless atmospheres. 

If clouds and color are independent, the binomial probability of having 7 or more cloudy dwarfs with 
$m_{\rm{F770W}}-m_{\rm{F1000W}}<0.03$~mag is 17\%, while the binomial probability of having at most one cloudless dwarf with similarly blue colors is essentially zero. This means that the joint probability of both independent results occurring at random is practically zero or, in other words, the separation between cloudy and cloudless atmospheres at $m_{\rm{F770W}}-m_{\rm{F1000W}}=0.03$~mag is significant ($\approx100\%$). 
This shows that CMDs including the MIRI F770W and F1000W filters allow us to anticipate whether an extrasolar atmosphere without mid-infrared spectra might have silicate clouds in its mid-infrared photosphere based solely on MIRI photometry. 

\subsubsection{Predictions from evolutionary models and atmospheric models}
\label{sec:models}

In this section we placed in context the trends of the late-M to T dwarfs in the CMDs with theoretical expectations. We consider the evolutionary models ATMO 2020 \citep{Phillips_2020} and Sonora Bobcat \citep{Marley_etal2021,Marley_etal2021b}, as well as sequences predicted by the Sonora Diamondback cloudy atmospheric models \citep{Morley_etal2024,Morley_etal2024b}.

The ATMO 2020 evolutionary models for substellar object (1--75~$M_{\rm Jup}$) cover  temperatures $T_{\rm eff} = 200 - 3000$~K (every 50--100~K) and gravity $\log(g) = 2.5-5.5 $ (every 0.5 dex). 
These models are composed of 3 sets of evolutionary sequences for cloudless objects, two grids assuming disequilibrium chemistry due to vertical mixing with eddy diffusion coefficient of $\log K_{zz}=4$ (weak mixing) and $\log K_{zz}=6$ (strong mixing), in addition to a grid assuming equilibrium chemistry ($\log K_{zz}=0$).

The Sonora Bobcat evolutionary models are also adequate for substellar objects (0.5--85~$M_{\rm Jup}$) and cover temperatures $T_{\rm eff} = 200 - 2400$~K and surface gravities $\log(g) = 2.5-5.5$. These models assume chemical equilibrium and cloudless atmospheres. 

The Sonora Diamondback atmospheric models provide synthetic spectra for cloudy substellar object atmospheres ($\approx1$--$84~M_{\rm Jup}$) with effective temperatures $T_{\rm eff} = 900 - 2400$~K and surface gravities $\log(g) = 3.5 - 5.5$. 
Clouds are parameterized using the sedimentation efficiency, $f_{\rm sed}$, which spans 1--8 as well as a cloud-free case, with smaller $f_{\rm sed}$ values corresponding to cloudier atmospheres.
These models assume equilibrium chemistry. Because no evolutionary models with JWST MIRI magnitudes currently exist for Sonora Diamondback or other cloudy atmospheric grids, we derived synthetic photometry directly from the Sonora Diamondback spectra for comparisons with our observations, as explained below.

As the native Sonora Diamondback grid samples $T_{\rm eff}$ in coarse steps of 100~K, we used the SEDA\footnote{https://seda.readthedocs.io/en/latest/} (Spectral Energy Distribution Analyzer) open-sorce Python package \citep[][Su\'arez et al. 2026, submitted]{Suarez_2021} to generate additional models with a finer temperature spacing of 20~K and derived JWST/MIRI synthetic photometry from the resulting synthetic spectra. This allowed us to populate the model sequences more densely and better trace the boundaries of the predicted model locus in the CMDs.

\begin{figure*}
    \centering
    Color-Magnitude Diagrams\par\medskip
	{\includegraphics[width=.44\linewidth]{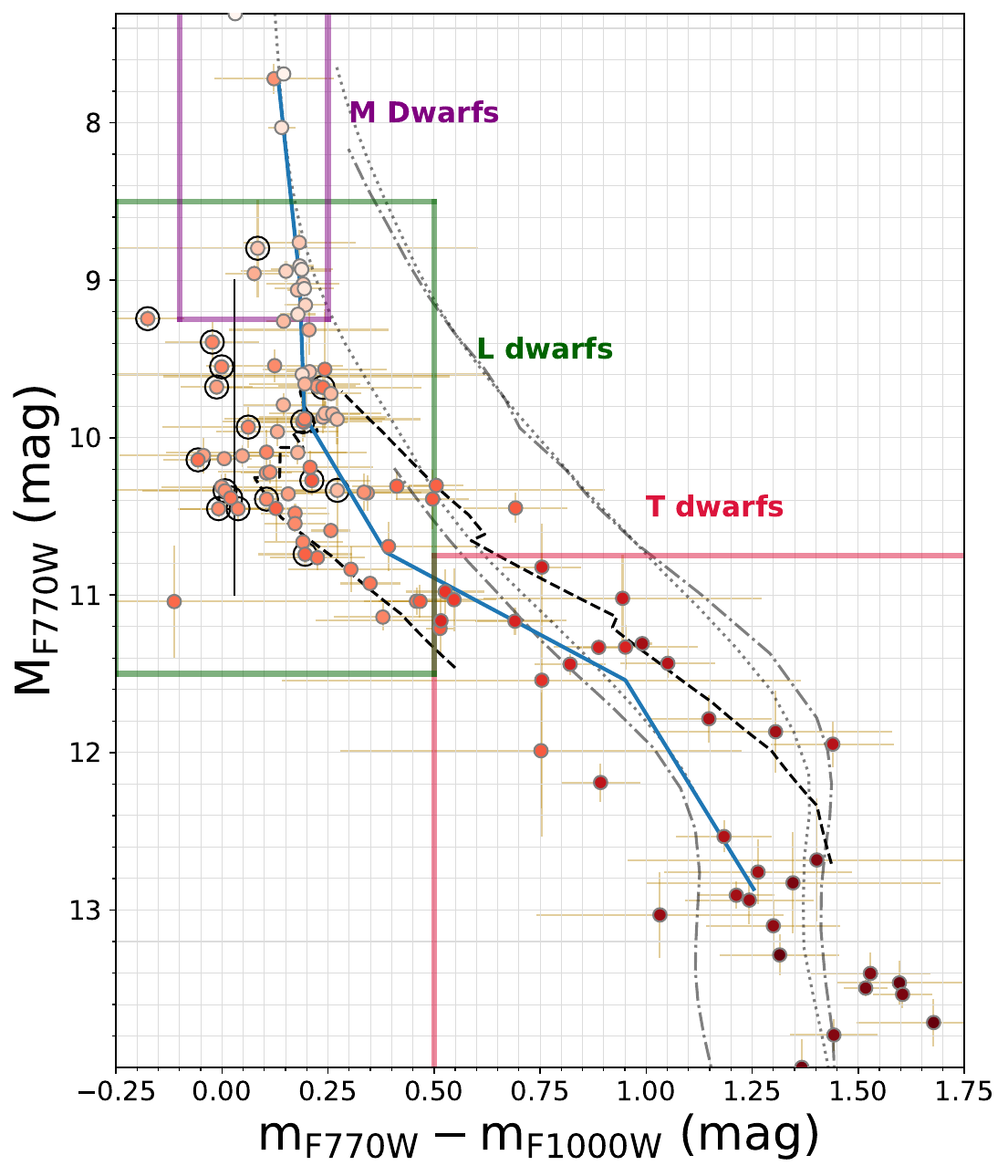}}
	{\includegraphics[width=.44\linewidth]{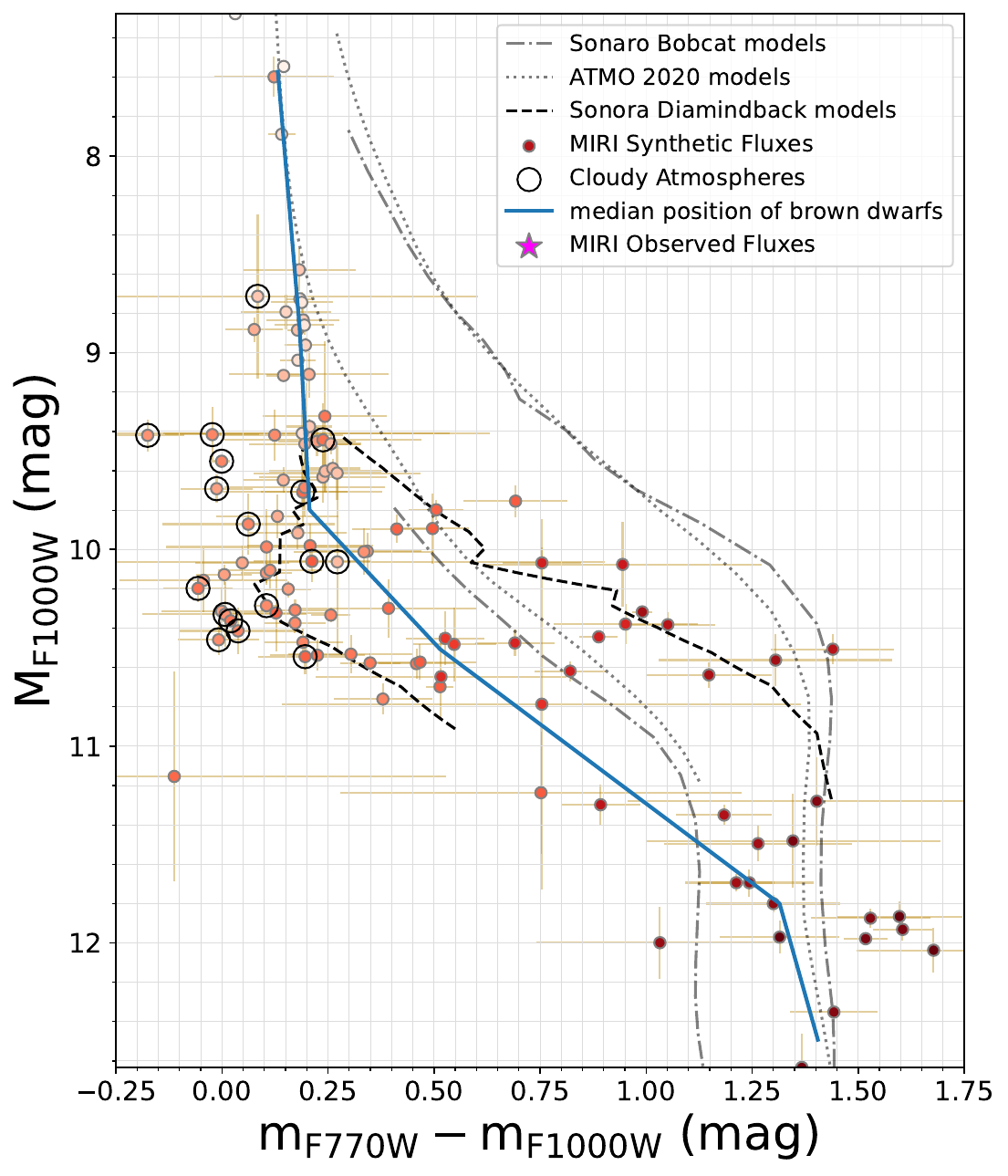}} \\
        {\includegraphics[width=.44\linewidth]{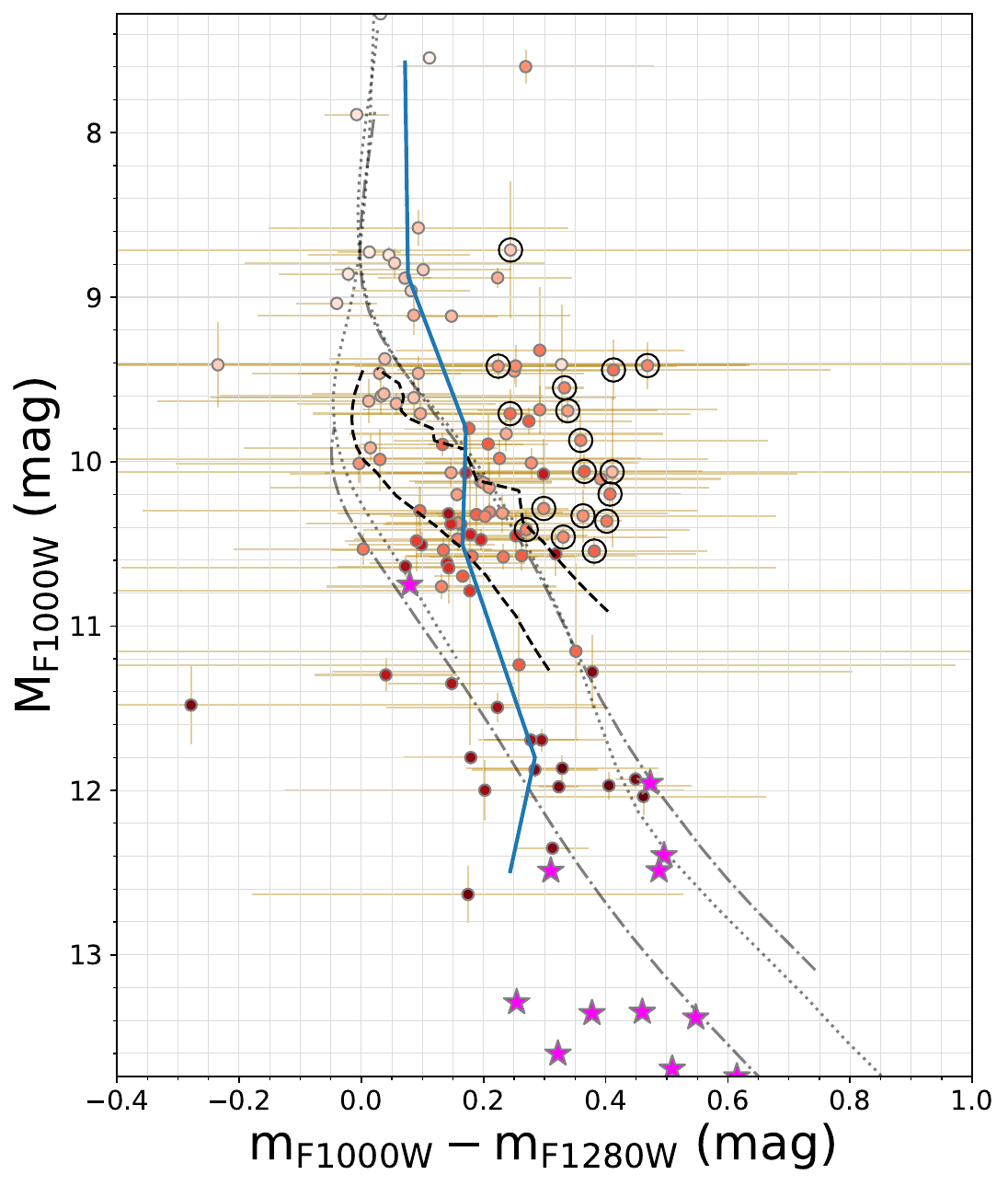}}
	{\includegraphics[width=.52\linewidth]{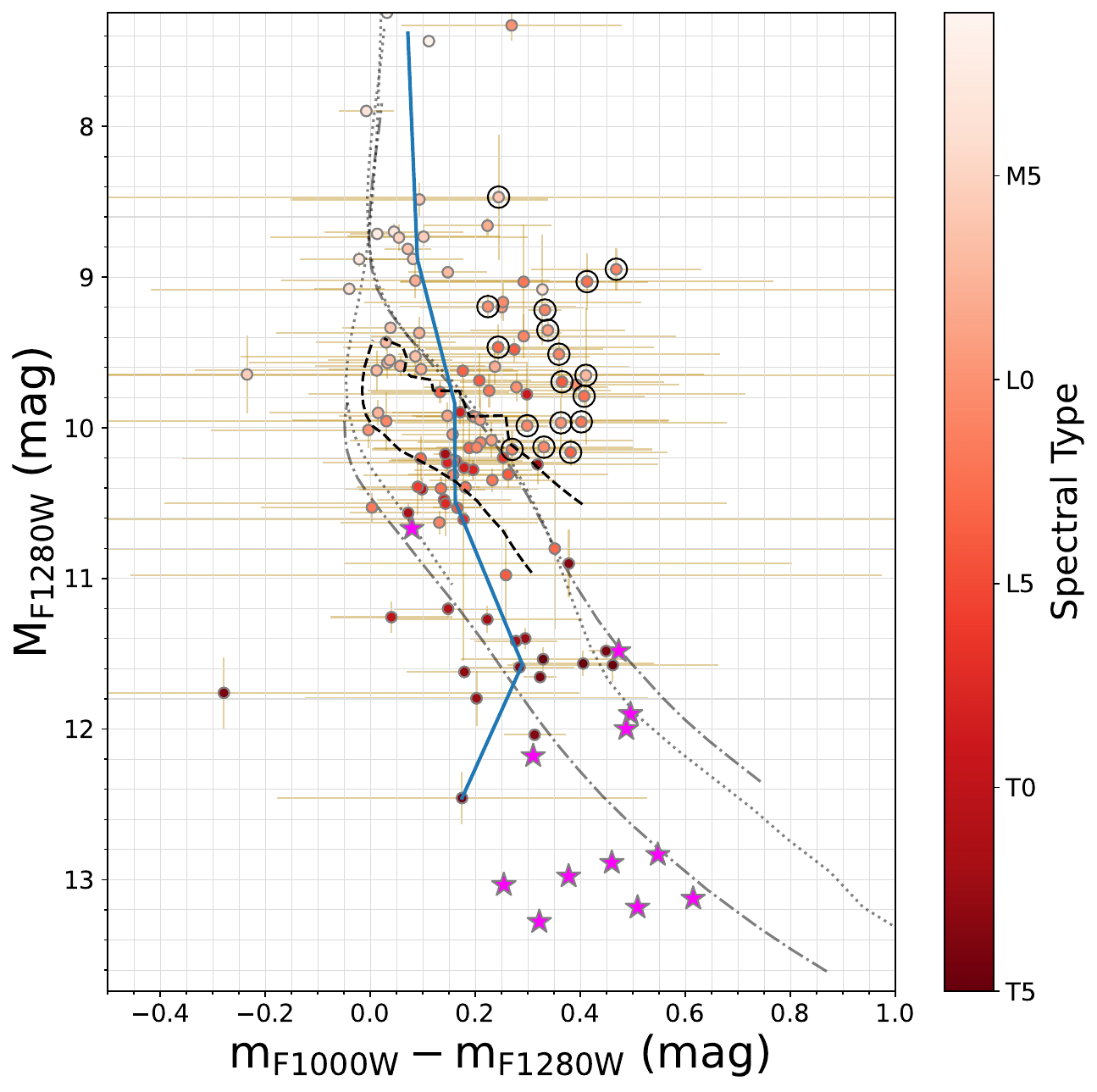}}
    \caption{$M_{\rm{F770W}}$ vs. $m_{\rm{F770W}}-m_{\rm{F1000W}}$ (top left panel), $M_{\rm{F1000W}}$ vs. $m_{\rm{F770W}}-m_{\rm{F1000W}}$  (top right panel), $M_{\rm{F1000W}}$ vs. $m_{\rm{F1000W}}-m_{\rm{F1280W}}$ (bottom left panel), and $M_{\rm{F1280W}}$ vs. $m_{\rm{F1000W}}-m_{\rm{F1280W}}$ (bottom right panel) CMDs for ultracool dwarfs using MIRI synthetic magnitudes. The blue curve represents the median position of dwarfs with similar spectral types (bins of 2 spectral subtypes). 
    Cloudy atmospheres are indicated by black open circles (see Section~\ref{sec:syn_color_mag}). 
    Dotted lines show 
    ATMO 2020 isochrones for 10 and 70~$M_{\rm Jup}$, dash-dotted lines show Sonora Bobcat sequences for $\log g$ of 4.0 and 5.5, and dashed lines show Sonora Diamondback predictions, delimited by models with $\log g = 5.5$, $Z=0$, $f_{\rm sed}=1$ (left) and $\log g = 4.0$, $Z=0$, no clouds (right)(see Section~\ref{sec:models}).   
    Spectral types are indicated by the colorbar in the bottom right panel. The bottom CMDs include observed MIRI magnitudes for late-T and Y dwarfs in JWST GO 2124 (fuschia stars; Section~\ref{sec:jwst_spectra}). 
    In the top left panel, the boxes roughly separate the M, L, and T spectral types (see Section~\ref{sec:syn_color_mag}) and the vertical line at 0.03~mag indicates the color threshold to distinguishing cloudy and cloudless objects.}
	\label{fig:CMDs}
\end{figure*}

Figure~\ref{fig:CMDs} shows CMDs comparing predictions from the cloudy and cloud-free models with the synthetic magnitudes derived from the Spitzer spectra.
We observe that evolutionary cloud-free models predict to some extent the location of the late-M and T dwarfs in our sample. However, 
these models fail to reproduce the position of most L dwarfs, which deviate significantly from the predicted trends, particularly in the $m_{\rm{F770W}}-m_{\rm{F1000W}}$ color. 
In this space, the majority of L dwarfs exhibit bluer colors compared to theoretical predictions. 
In contrast, when considering $m_{\rm{F1000W}}-m_{\rm{F1280W}}$ colors, only cloudy L dwarfs tend to deviate from the prediction, exhibiting redder colors.

We expect that cloudy objects in Figure~\ref{fig:CMDs} deviate from the predictions of the ATMO 2020 and Sonora Bobcat evolutionary models because these models do not consider clouds. As a result, they cannot account for silicate absorption and thus fail to reproduce the blue $m_{\rm{F770W}}-m_{\rm{F1000W}}$ colors and the red $m_{\rm{F1000W}}-m_{\rm{F1280W}}$ colors observed in L dwarfs from the Spitzer IRS sample. 

The cloudy Sonora Diamondback models better reproduce the location of L dwarfs in $m_{\rm{F770W}}-m_{\rm{F1000W}}$ color by predicting bluer colors than the cloud-free models. However, in the $m_{\rm{F1000W}}-m_{\rm{F1280W}}$ color these models face similar challenges as the cloudless models, predicting significantly bluer colors than observed
for most cloudy L dwarfs. 
This suggests that the Sonora Diamondback models more accurately capture the indirect effects of silicate clouds on the 6.25~$\mu$m water absorption feature but struggle to reproduce the $\sim$9~$\mu$m silicate absorption feature, which is a known limitation of self-consistent cloudy atmospheric models \citep[e.g.,][]{Morley_etal2024,Petrus_etal2024,Campos_Estrada_etal2025}. 
Consequently, more sophisticated approaches, such as modified self-consistent models \citep[e.g.,][]{Luna_Morley2021} and atmospheric retrievals \citep[e.g.,][]{Hoch_etal2025,Molliere_etal2025}, have been developed to better reproduce the silicate absorption feature. 
For T dwarfs, the Sonora Diamondback model predictions for the MIRI colors are consistent with those from the cloud-free models and align with the observed data.

\begin{figure}[ht]
    \includegraphics[width=1\linewidth]{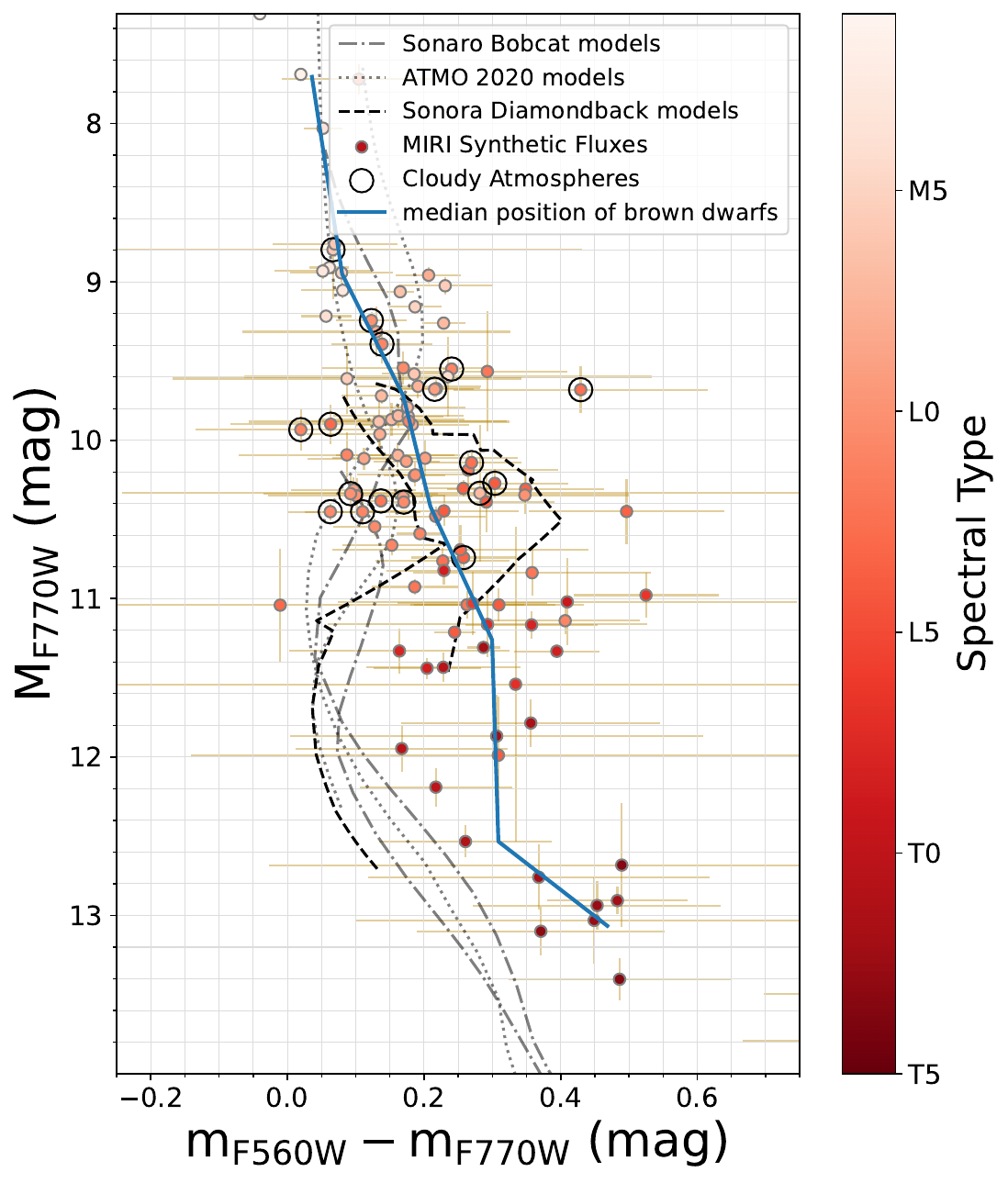}
	\caption{$M_{\rm{F770W}}$ vs. $m_{\rm{F560W}}-m_{\rm{F770W}}$ CMD for late-M to T dwarfs. Elements in the figure are the same as in Figure~\ref{fig:CMDs}.}
    \label{fig:CMD2}
\end{figure}

The evolutionary model sequences converge in the $M_{\rm{F770W}}$ vs. $m_{\rm{F560W}}-m_{\rm{F770W}}$ CMD shown in Figure~\ref{fig:CMD2}, although the cloudy models are slightly shifted towards redder colors for L dwarfs, providing a better match to the overall trend of these objects. 
This improved agreement supports the conclusion from the previous paragraph that cloudy atmospheric models adequately reproduce the indirect impact of silicate clouds on the 6.25~$\mu$m water feature probed by both the F560W and F770W filters.

For the warmest objects in our sample, the models in general explain their position in the $M_{\rm{F770W}}$ vs. $m_{\rm{F560W}}-m_{\rm{F770W}}$ CMD but struggle to reproduce the cooler ones. 
In particular, T dwarfs have redder $m_{\rm{F560W}}-m_{\rm{F770W}}$ colors than predicted. 
We speculate it may be due to a smaller relative difference between the predicted depths of the 6.25~$\mu$m water and the $7.65~\mu$m methane absorptions compared to the observed one in the Spitzer IRS spectra.

The mid-infrared water and methane features strengthen with spectral type \citep{Cushing_2006,Suarez_2022}, so both $m_{\rm{F560W}}$ and $m_{\rm{F770W}}$ become fainter with later spectral types. This behavior likely leads to the predicted narrow $m_{\rm{F560W}}-m_{\rm{F770W}}$ color range across the full late-M to T sample and results in more stacked evolutionary sequences in the CMD defined by F560W and F770W (Figure~\ref{fig:CMD2}) compared to the CMDs in Figure~\ref{fig:CMDs}.

\subsection{Synthetic Color-Color Diagrams}
\label{sec:syn_color_color}

\begin{figure}[ht]
    \includegraphics[width=1\linewidth]{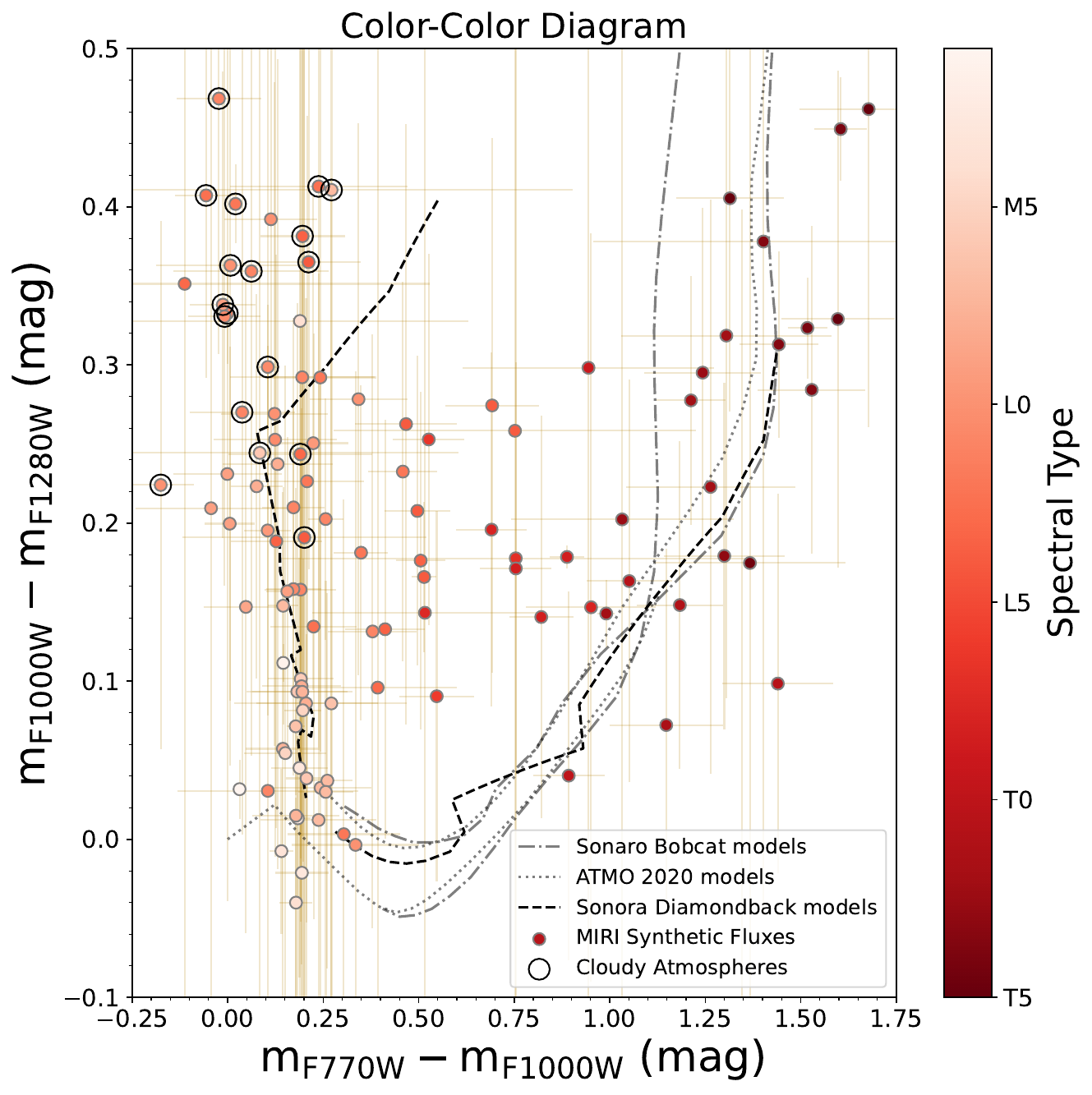}
	\caption{$m_{\rm{F1000W}}-m_{\rm{F1280W}}$ vs. $m_{\rm{F770W}}-m_{\rm{F1000W}}$ CCD for the late-M to T dwarfs with Spitzer IRS spectra. Elements in the figure are the same as in Figure~\ref{fig:CMDs}.}
    \label{fig:color_color_models}
\end{figure}

In addition to the CMDs in Figures~\ref{fig:CMDs} and \ref{fig:CMD2}, we examined the 
$m_{\rm{F1000W}}-m_{\rm{F1280W}}$ vs. $m_{\rm{F770W}}-m_{\rm{F1000W}}$ CCD shown in Figure~\ref{fig:color_color_models}. 
The location of the objects in this space is quite uncertain, mainly due to the large uncertainties of the $m_{\rm{F1000W}}-m_{\rm{F1280W}}$ colors caused by the low signal of the IRS spectra at long wavelengths. 
Nevertheless, general trends can still be identified.

We observe in Figure~\ref{fig:color_color_models} that most warm brown dwarfs exhibit relatively blue $m_{\rm{F770W}}-m_{\rm{F1000W}}$ colors while spanning a wide range of $m_{\rm{F1000W}}-m_{\rm{F1280W}}$ colors ($\approx0-0.5$~mag). 
In contrast, the coolest objects are generally redder in both colors, although the scatter in the data is higher.

Cloud-free models form a narrow sequence that trends toward redder colors, roughly following the behavior of T dwarfs but. However, as expected, they fail to reproduce the position of L dwarfs, particularly the reddest $m_{\rm{F1000W}}-m_{\rm{F1280W}}$ colors observed in objects with cloudy atmospheres. 
In comparison, prediction by cloudy models successfully reproduce the overall trend of redder $m_{\rm{F1000W}}-m_{\rm{F1280W}}$ colors for later spectral types at nearly constant $m_{\rm{F770W}}-m_{\rm{F1000W}}$ color ($\sim$0.2~mag), as observed for L dwarfs.



\subsection{Extension to Y-dwarf using JWST MIRI Observations}
\label{sec:jwst_ydwarfs}

To explore the extension of the trends in mid-infrared CMDs into cooler temperatures or later spectral types, we consider JWST MIRI observations of T/Y transition dwarfs. 
We plotted in the bottom panel CMDs in Figure~\ref{fig:CMDs} the 12 T/Y transition dwarfs with MIRI F1000W, F1280W, and F1800W photometry in JWST Cycle 1 GO 2124 program (Section~\ref{sec:jwst_spectra}). 
Within the data dispersion, these T/Y dwarfs appear to follow the reddening trend towards fainter magnitudes observed for the L and T dwarfs with synthetic magnitudes from Spitzer IRS spectra. 
We also note a significant scatter in the location of the T/Y dwarfs that cannot be explained by photometric uncertainties. 
As observed in Figure~\ref{fig:plot_subset}, the dominant spectral feature in T/Y transition objects in the wavelengths covered by the F1000W and F1280W filters ($\approx$8.8--14.2~$\mu$m) is an ammonia absorption feature centered at 10.5~$\mu$m. 
This raises the possibility that the dispersion observed in the T/Y dwarfs may reflect a diversity of such 10.5~$\mu$m ammonia feature \citep[e.g.;][]{Beiler_etal2024}, whose depth can be significantly affected by atmospheric heating \citep{Suarez_2025}, among other factors.

We observe in Figure~\ref{fig:CMDs} (bottom panel) that models face challenges in explaining the location of the T/Y transition dwarfs. They predict redder $m_{\rm{F1000W}}-m_{\rm{F1280W}}$ colors compared to the JWST MIRI observations of the coldest objects in our analysis. 
The cause of this issue remains uncertain. 
However, we note that the 10.5~$\mu$m ammonia absorption feature affects $m_{\rm{F1000W}}$ but not $m_{\rm{F1280W}}$ (Figure~\ref{fig:plot_subset}).
If this feature is stronger in the models, then $m_{\rm{F1000W}}$ would be fainter, producing redder $m_{\rm{F1000W}}-m_{\rm{F1280W}}$ colors than observed. 
This suggests that the models may overestimate the strength of the 10.5~$\mu$m ammonia absorption feature for brown dwarfs at the T/Y transition. 
Future forward-modeling analyses of mid-infrared spectra for cold brown dwarfs could test the accuracy of the modeled ammonia absorption and help resolve this discrepancy.

\begin{figure}[ht]
    \includegraphics[width=1\linewidth]{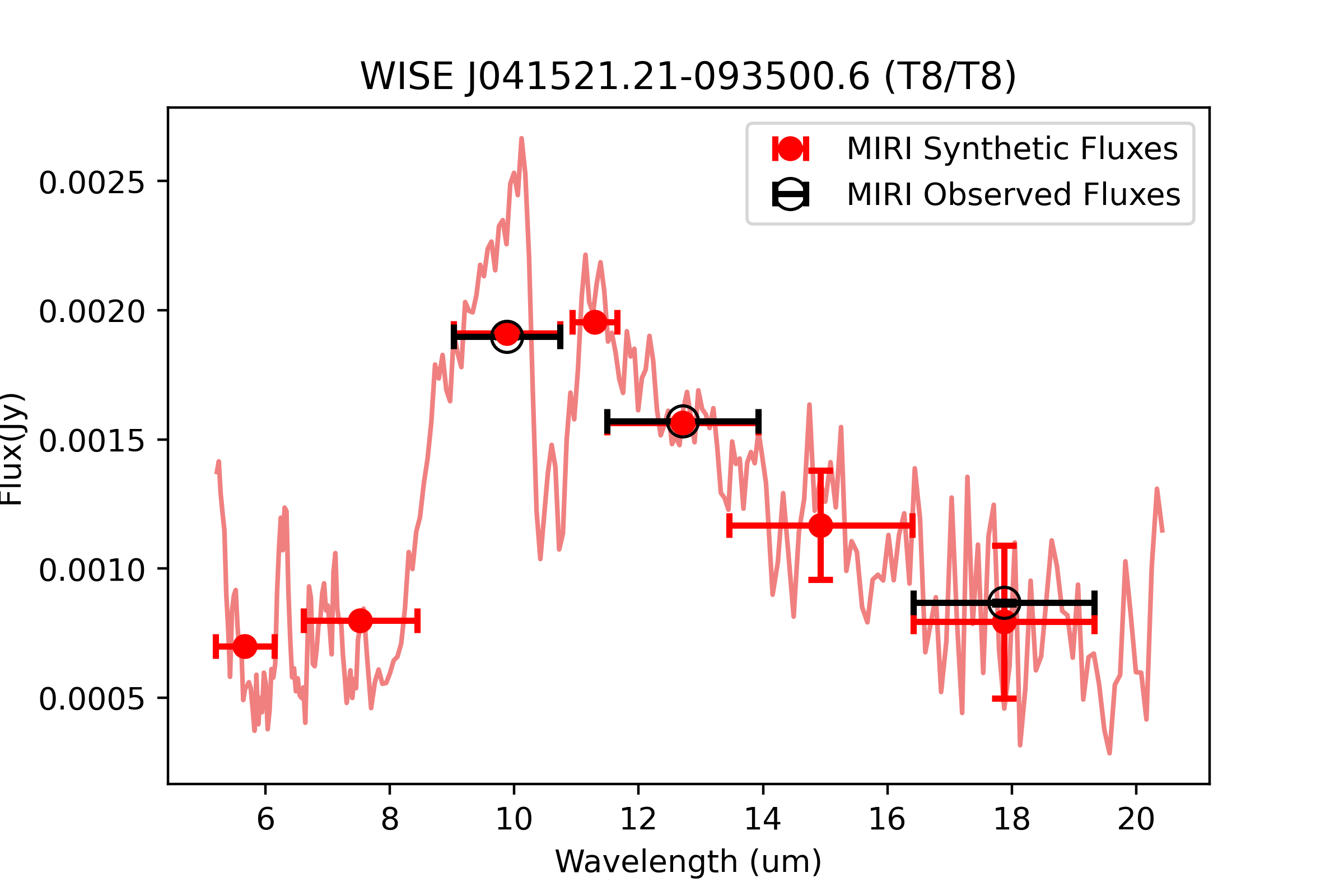}
	\caption{Spitzer IRS spectrum of WISE J041521.21-093500.6 along with synthetic MIRI F560W, F770W, F1000W, F1280W, F1500W, and F1800W fluxes in red and observed MIRI F1000W, F1280W, and F1800W fluxes in black.}
    \label{fig:jwst_comparison}
\end{figure}

The JWST MIRI observations and the synthetic MIRI photometry from Spitzer IRS spectra also allow us to verify the accuracy of the photometry automatically provided by the JWST calibration pipeline. 
The T8 dwarf WISE J041521.21-093500.6 has both MIRI observations and MIRI synthetic photometry. 
Figure~\ref{fig:jwst_comparison} shows the Spitzer IRS spectrum for this dwarf along with the derived and observed synthetic fluxes. 
The observed F1000W, F1280W, and F1800W fluxes are in great agreement with the derived synthetic fluxes.



\section{Summary}
\label{sec:summary}

We presented mid-infrared JWST MIRI CMDs and a CCD to identify the location of objects with silicate clouds in their atmospheres and tested how well model sequences predict the evolution of the warmest to the coldest brown dwarfs. For this, we derived synthetic photometry for the JWST MIRI F560W (4.9--6.489 $\mu$m), F770W (6.5--8.8 $\mu$m), F1000W (8.8--11.190 $\mu$m), and F1280W (11.3--14.3 $\mu$m) filters using Spitzer IRS spectra of the 113 late-M to T dwarfs in \cite{Suarez_2022} and incorporated MIRI imaging data acquired with the F1000W, F1280W, and F1800W filters for T/Y transition dwarfs.

We found that the CMDs involving the F770W and F1000W filters separate well cloudy objects from cloudless dwarfs (top panel in Figure~\ref{fig:CMDs}). 
In particular, L dwarfs or objects with $M_{\rm{F770W}}\approx$9.2--10.5~mag having $m_{\rm{F770W}}-m_{\rm{F1000W}}$ colors bluer than 0.03~mag are seven times more likely to have cloudy atmospheres. 
The estimated contamination of cloudless atmospheres with such blue colors is only 12\%. Regarding the recovery of known cloudy objects,  
we found that 70\% of the cloudy atmospheres meet the color condition, while the 30\% remaining cloudy dwarfs have redder colors. 
Thus, the $m_{\rm{F770W}}-m_{\rm{F1000W}}<0.03$~mag color condition can be used to select new potential cloudy atmospheres with only photometric information for further follow-up spectroscopy to study the properties and composition of silicate cloud in substellar atmospheres.

CMDs involving the F1000W and F1280W filters (bottom panel in Figure~\ref{fig:CMDs}) and the CCD with these filters plus F770W (Figures~\ref{fig:color_color_models}) also appear to be effective at distinguishing cloudy atmospheres. 
However, the results are less conclusive due to larger uncertainties in the positions of the objects in the diagrams, driven by the lower S/N of the Spitzer IRS spectra at the longest wavelengths, which mainly affects the F1280W synthetic photometry.

The trends of brown dwarfs in the mid-infrared CMDs and CCD cannot be fully explained by either the ATMO 2020 and Sonora Bobcat evolutionary models or the predictions by the Sonora Diamondback atmospheric models. 
The main discrepancies arise for L-type brown dwarfs, particularly those with cloudy atmospheres. 
Cloudy models improve the prediction for these objects when considering MIRI filters that probe the 6.25~$\mu$m water spectral feature, but discrepancies persist for other MIRI filters. 
This suggests that the Sonora Diamondback cloudy atmospheric models successfully capture the impact of silicate clouds on the depth of the water absorption feature but fail to reproduce the $\sim9~\mu$m silicate absorption observed in most L dwarfs (Figure~\ref{fig:plot_subset}).

The use of MIRI CMDs and CCDs in this study represents a novel technique for identifying new objects with silicate clouds in their atmospheres in both new and archival MIRI imaging data, particularly in sky regions of high object density, such as young stellar associations, for targeted spectroscopic follow-up.

The JWST spectrum presented in this article were obtained from the Mikulski Archive for Space Telescopes (MAST) at the Space Telescope Science Institute. The specific observations analyzed can be accessed via \dataset[doi: 10.17909/k7qh-am76]{https://doi.org/10.17909/k7qh-am76}.

\section*{Acknowledgements}
JL acknowledges support from the New York Space Grant Consortium NASA under Grant 80NSSC20M0096. GS and JKF acknowledge support from NASA award 80NSSC22K0142 and STSCI award JWST-GO-02124.001-A. SM acknowledges support from the Canadian Space Agency via grant 23JWGO2A12. JMV acknowledges from a Royal Society - Research Ireland University Research Fellowship (URF/1/221932) and the European Union through the Exo-PEA ERC project (grant number 101164652). Views and opinions expressed are however those of the author(s) only and do not necessarily reflect those of the European Union or the European Research Council Executive Agency. Neither the European Union nor the granting authority can be held responsible for them.

\bibliography{sample631}{}
\bibliographystyle{aasjournal}

\end{document}